\pdfoutput=1

\documentclass{iau} 
\usepackage{graphicx}

\title[Magnetic Activity in the Galactic Center by Global MHD Simulation] 
      {Vertical flows and structures 
      excited by magnetic activity 
      in the Galactic center region}

\author[Kakiuchi et al.]   
       {Kensuke Kakiuchi$^1$,Takeru K. Suzuki$^{1,2}$, Yasuo Fukui$^1$, \\
	Kazufumi Torii$^{1,3}$, Mami Machida${^4}$, \and Ryoji Matsumoto$^{5}$
          }

       \affiliation{$^1$Dept. of Physics, Nagoya University, Furo-cho, Chikusa, 
       Nagoya, Aichi, 464-8602, Japan \\email: {\tt kakiuchi@nagoya-u.jp} \\[\affilskip]
   $^2$School of Arts \& Sciences, The University of Tokyo, \\
         3-8-1, Komaba, Meguro, Tokyo, 153-8902, Japan \\ 
  $^3$Nobeyama Radio Observatory, National Astronomical Observatory of Japan, 462-2,
   Nobeyama, Minamimaki, Minamisaku, Nagano, 384-1305, Japan\\
  $^4$Dept. of Physics, Faculty of Sciences, Kyushu University, 
  6-10-1 Hakozaki, Higashi-ku, Fukuoka 812-8581, Japan \\
  $^5$Dept. of Physics, Graduate School of Science, Chiba University,
  1-33 Yayoi-cho, Inage-ku, Chiba 263-8522, Japan}

\pubyear{2016}
\volume{322}  
\jname{Title of your IAU Symposium}
\editors{A.C. Editor, B.D. Editor \& C.E. Editor, eds.}
\begin{document}

\maketitle

\begin{abstract}
Various observations show peculiar features in the Galactic center region, such as loops and filamentary structure. It is still unclear how such characteristic features are formed. Magnetic field is believed to play very important roles in the dynamics of gas in the Galaxy Center. \cite[Suzuki et al.(2015)]{suzuki15a} performed a global magneto-hydrodynamical simulation focusing on the Galactic Center with an axisymmetric gravitational potential and claimed that non-radial motion is excited by magnetic activity. We further analyzed their simulation data and found that vertical motion is also excited by magnetic activity. In particular, fast down flows with speed of $\sim$100 km/s are triggered near the footpoint of magnetic loops that are buoyantly risen by Parker instability. These downward flows are accelerated by the vertical component of the gravity, falling along inclined field lines. As a result, the azimuthal and radial components of the velocity are also excited, which are observed as high velocity features in a simulated position-velocity diagram. Depending on the viewing angle, these fast flows will show a huge variety of characteristic features in the position-velocity diagram.

\end{abstract}

\section{Introduction \& Method}
Magnetic field is regarded to play an important role in the gas in the Galactic center region. 
Recent polarization observation reported that a dark infrared cloud possesses
magnetic field with strength $\sim$ a few mG.
Such a strong magnetic field is expected to affect the gas dynamics near the Galactic center. 
\cite[Suzuki et al. (2015)]{suzuki15a} performed a global MHD simulation focusing on the Galactic bulge region under the axisymmetric Galactic potential \cite[(Miyamoto \& Nagai 1975)]{miyamoto75a} by modifying the three dimensional MHD simulation code originally developed for protoplanetary disks \cite[(Suzuki  \& Inutsuka 2009; 2014; Suzuki et al.2010)]{suzuki09a,suzuki14a,suzuki10a}. 
In this proceedings paper, we investigated vertical flows and magnetic structure obtained from the numerical data, which were not analyzed in detail in \cite[Suzuki et al. (2015)]{suzuki15a}. 


              

\section{Results}
We analyzed flows and magnetic fields obtained from the MHD simulation by \cite[Suzuki et al. (2015)]{suzuki15a}. 
Figure 1 shows trajectories of fluid elements from $t=399.5$ to $405.5$ Myr located in 0.3-0.4 kpc. 
These fluid elements rotate in the clockwise direction with speed $\approx 200$ km s$^{-1}$.
In addition to the rotation, some fluid elements show fast vertical and/or radial motion with speed $\sim 10-100$ km s$^{-1}$.
In Figure 2 we present the velocity and magnetic fields near a fast-downflow region.  
The downflow is accelerated by the vertical component of the gravity along the inclined field lines, which we call a magnetic sliding slope.
The magnetic sliding slope is a part of a magnetic loop that is buoyantly lifted up by Paker instability \cite[(Parker 1966)]{parker66a}.
Complex loop structures are observed near the Galactic center \cite[(Fukui et al. 2006; Machida et al. 2009; Torii et al. 2010a)]{fukui06a,machida09a,torii10b}.
Our result indicates that downflows can be associated with these loops as discussed in \cite[Torii et al. (2010b)]{torii10a}.


\begin{figure}[h]
   \centering
   \includegraphics[clip,width=4.2in]{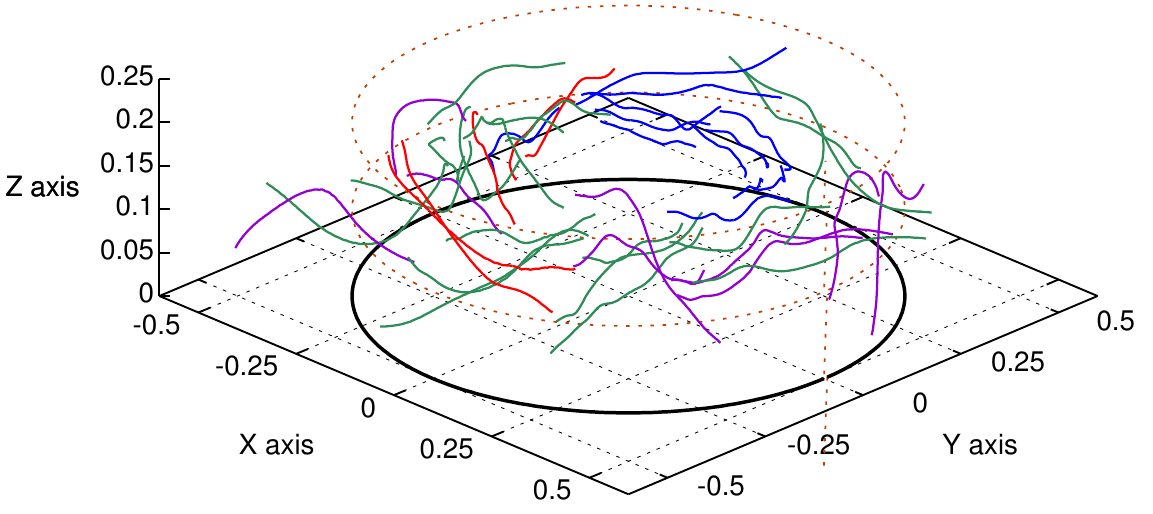}
  \caption{\small
    Trajectories of fluid elements from $t=399.5$ to $405.5$ Myr.
    The unit of the axes is kpc. 
    Flow patterns are classified by colours, red (monotonic upflows)
    green (monotonic downflows), purple (combination of upflows and downflows),
    and blue (no apparent up/downflow).    
   }
   \label{fig_traject}
\end{figure}

\begin{figure}[ht]
   \centering{
   \includegraphics[clip,width=4.2in]{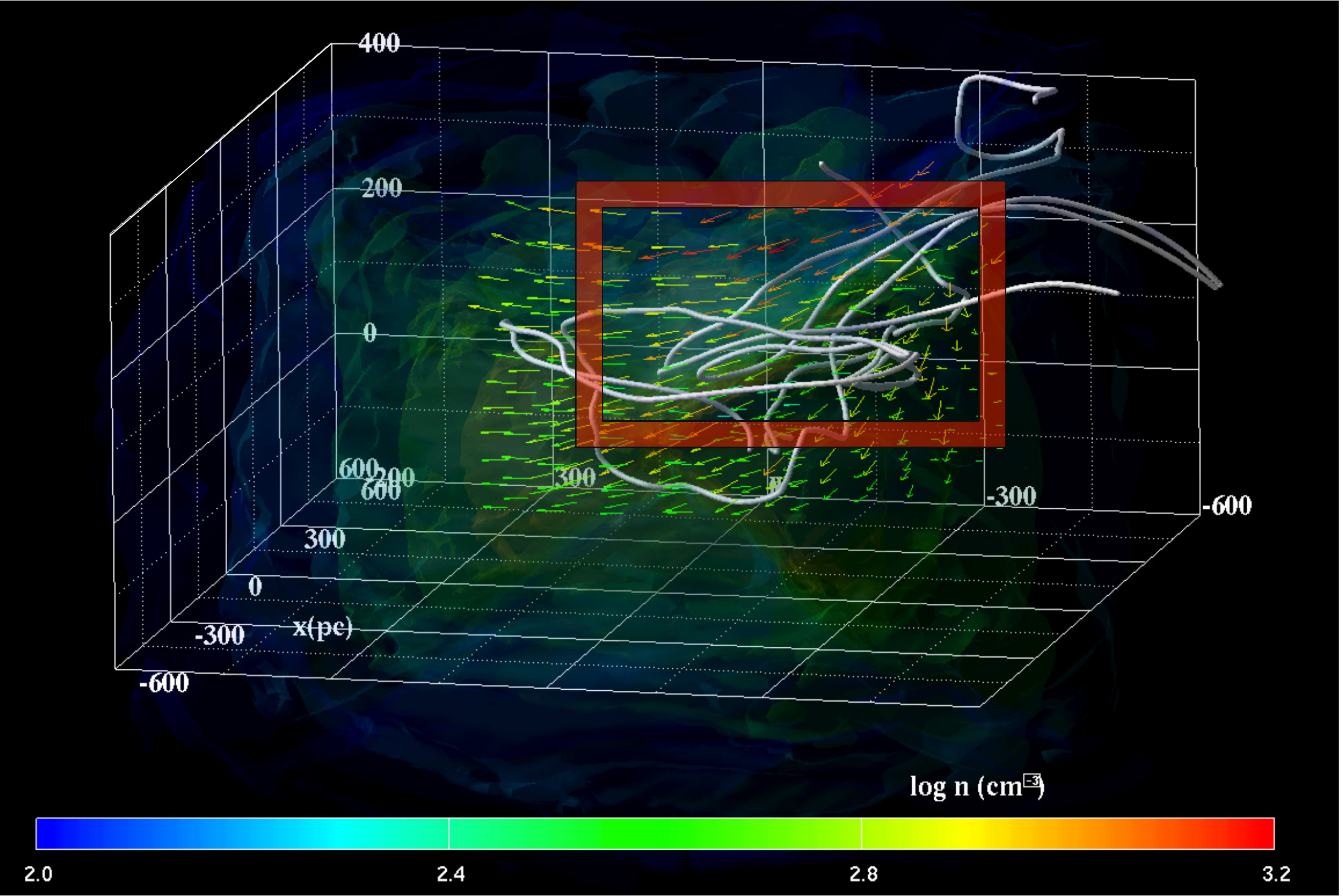}}
  \caption{\small 
    Magnetic field lines (white lines) and velocity field (arrows) near a fast-downflow region.
  }
   \label{fig_magline}
\end{figure}

\section{Conclusion}
Our analyses of the numerical data obtained from the global MHD simulation by \cite[Suzuki et al. (2015)]{suzuki15a} showed that the magnetic activity excites vertical magnetic structure and vertical flows.
Those vertical flows will be possible observed as a peculiar feature in a
position -- velocity diagram; we plan to inspect this in the forthcoming paper
(Kakiuchi et al. 2017, in preparation).


\nocite{*}

\bibliography{IAU322}

%
%
%
%
%
%
%

\end{document}